\documentclass[12pt]{article}
\usepackage[dvipsnames]{xcolor}
\usepackage{color}
\usepackage[english]{babel}
\usepackage{amsmath}
\usepackage{amsfonts}
\usepackage{amssymb}
\usepackage{graphicx}
\def\ai{\'{\i}}
\def\lp{\left(}
\def\rp{\right)}
\def\lb{\left[}
\def\rb{\right]}
\def\be{\begin{equation}}
\def\ee{\end{equation}}
\def\Om{\Omega}
\def\om{\omega}
\begin{document}
\title{\vspace{-1.5cm} \bf Further considerations about the traversability of thin-shell wormholes}
\author{E. Rub\ai n de Celis\footnote{e-mail: erdec@df.uba.ar} and C. Simeone\footnote{e-mail: csimeone@df.uba.ar}\\
{\footnotesize Departamento de F\ai sica, Facultad de Ciencias Exactas y Naturales, Universidad de}\\
{\footnotesize Buenos Aires and IFIBA, CONICET, Ciudad Universitaria, Buenos Aires 1428, Argentina.}}
\date{\small \today}

\maketitle
\vspace{0.6cm} 
\begin{abstract}

Traversability in relation with tides in thin-shell wormholes is revisited to investigate the possibility of improving recently noted restrictive conditions for a safe travel across a wormhole throat. We consider wormholes mathematically constructed starting from background geometries which are solutions of scalar-tensor theories as dilaton gravity and Brans--Dicke gravity. The advantages of working within such frameworks are studied by examining the dependence of the extrinsic curvature and tides at the throat with the parameters determining the departure from pure relativity; the associated behaviour of tides in the smooth regions of the geometries is also analyzed. Other related but different approaches are briefly discussed in the appendices.

\vspace{1cm} 

\noindent 
PACS number(s): 4.20.-q, 04.20.Gz, 04.40.Jb\\


\end{abstract}

\vspace{1cm}

%\end{document}

\section{Introduction}

The recent study in \cite{nos21} shows that the main difficulty with tides across the throat of thin-shell wormholes does not come from the contribution of the smooth parts of the geometries: for points along both the radial or an angular direction, the corresponding effects are finite and proportional to the spatial separation of these points, and could be controlled by an appropriate choice of the parameters defining the metrics at each side of the shell placed at the throat radius (see also Ref. \cite{martin}). Instead, the contribution coming from the jump in the extrinsic curvature across the shell presents some subtleties. For the radial tides this contribution is fixed and does not vanish for infinitely close points; hence the quotient of the relative acceleration and the separation between two points at different sides of the throat $\Delta a/\Delta x$ would, in principle, diverge when $\Delta x\to 0$. For the angular tides the problem is different: though the contribution of the curvature jump is proportional to the angular separation, it appears regulated by the traveling time across the shell, which, in a plain formal approach, is infinitely short.  These results would seem to rule out the possibility of a safe passage through such configurations.  However, in our previous work \cite{nos21} we briefly introduced an alternative point of view, in the spirit of taking the formal results as a first approximation to more realistic situations in which shells have a non vanishing --though still very little-- thickness. Within this understanding of the problem, here we will investigate the conditions to, at least, reduce the jump of the components of the extrinsic curvature across the shell; when such reduction is possible, we will also evaluate the effects of these conditions on the curvature of the smooth regions. In particular, we will explore the possible advantages, if any exist, of considering wormhole constructions in the framework of gravity theories beyond relativity, in particular those including a scalar field as the so-called dilaton gravity and Brans--Dicke gravity. We also include appendices dealing with different simple lines of analysis for other examples within the relativistic framework.

\section{General aspects of tides in thin-shell wormholes}

We restrict our analysis to static thin-shell wormholes constructed by pasting together two copies of the same geometry at the throat hypersurface located at $r=r_0$; the   line elements are
\be\label{metric1}
ds^2 = g^{\pm}_{00}\, dt^2 + g^{\pm}_{rr}\, dr_{\pm}^2 + g^{\pm}_{\zeta\zeta}\, d\zeta^2 + g^{\pm}_{\varphi\varphi}\, d\varphi^2 \,,
\ee
where the sign $\pm$ refers to each side of the throat with perpendicular radial coordinates $r_{\pm} \geqslant r_0$, respectively.  The $\zeta$ coordinate represents the polar
coordinate $\theta \in [0,\pi)$
in the case of spherical symmetry, or the axial coordinate 
$z \in \mathbb{R}$
when we consider a cylindrical geometry. In the second case the metric elements depend  on $r_{\pm}$; in the first one, the coefficient $g^{\pm}_{\varphi\varphi}$ depends also on $\theta$. As usual, $t\in\mathbb{R}$ and $\varphi \in [0,2\pi)$. The normal coordinate associated to the radial direction is defined as
\be
d\eta = \pm \sqrt{g^{\pm}_{rr}} \, dr_{\pm} \,,
\ee
i.e., ($\pm$)$\eta$ measures the perpendicular proper distance at the vicinities of the throat located at $\eta = 0$, for a static observer; the unit normal vector to the shell is correspondingly defined as $n_{\mu}  = \partial_{\mu} \eta \,$, pointing from $-$ to $+$. 

Tides acting on a body traversing a wormhole are best described by the covariant relative acceleration usually expressed  as \cite{book,grav}
\be\label{A}
(\Delta a)^\mu =
- g^{\rho\mu}{R}_{\rho\alpha\nu\beta} V^\alpha (\Delta x)^\nu V^\beta.
\ee   
where ${R^\mu}_{\alpha\nu\beta}$ is the Riemann tensor, $V^\mu$ is the four-velocity, and $(\Delta x)^\mu$ is  a vector which stands for the small separation of two points in spacetime.

The components of the Riemann tensor can be put in the form
 \be
R_{\mu\alpha\nu\beta} 
=  \Theta(-\eta) R^{-}_{\mu\alpha\nu\beta} + \Theta(\eta) R^{+}_{\mu\alpha\nu\beta} 
- \delta(\eta) 
\left[\kappa_{\alpha\beta} \, n_\mu n_\nu+\kappa_{\mu\nu} \, n_\alpha n_\beta-\kappa_{\alpha\nu} \, n_\mu n_\beta-\kappa_{\mu\beta} \, n_\alpha n_\nu\right] 
\ee
where $R^{\mp}_{\mu\alpha\nu\beta}$ is the smooth tensor at each side of the shell and
\be
\kappa_{\alpha\beta} = \frac{1}{2} \left(\frac{\partial g^+_{\alpha\beta}}{\partial \eta} - \frac{\partial g^-_{\alpha\beta}}{\partial\eta} \right)\Big|_{r_0}
\ee
is the jump in the extrinsic curvature tensor at the shell. 

Given the symmetry of the problem, the components of the Riemann tensor which are relevant  for our analysis are  
\begin{eqnarray}
{R^{r0}}_{r0} & = & -\,\frac{1}{4g_{rr}g_{00}}\left\{ 2g_{00,rr}-g_{00,r}\frac{\lp g_{00}g_{rr}\rp_{,r}}{g_{00}g_{rr}}\right\},\\
{R^{\varphi 0}}_{\varphi 0} & = & -\,\frac{g_{\varphi\varphi,r}g_{00,r}}{4g_{rr}g_{00}g_{\varphi\varphi}},\\
{R^{\varphi r}}_{\varphi r}& = & {R^{\varphi 0}}_{\varphi 0} -\frac{1}{4g_{rr}g_{\varphi\varphi}}\left\{ 2g_{\varphi\varphi,rr}-g_{\varphi\varphi,r}\frac{\lp g_{00}g_{\varphi\varphi}g_{rr}\rp_{,r}}{g_{00}g_{\varphi\varphi}g_{rr}}\right\}
\end{eqnarray}
and analogous expressions substituting $\varphi$ by $z$ for cylindrically symmetric problems. 

With these definitions, we can recall the result of \cite{nos21} for the radial tide
of a radially moving object
\be\label{A}
\Delta a=\left[  {{R^{r0}}_{r0}}^{-} +{{R^{r0}}_{r0}}^{+}\right]_{r_0} 
\frac{\Delta \tilde\eta}{2} 
+ \mathcal{O}(\Delta \tilde\eta^2) 
-
{\kappa^{0}}_{0},
\ee
where ${\kappa^{0}}_0$ is the jump of the component ${K^{0}}_0$ of the extrinsic curvature across the infinitely thin shell, and $\Delta\tilde\eta$ is the proper radial separation between points of the object traversing the throat; this reduces to $\Delta\eta$ for a rest body. The smooth regions contribute with a term proportional to $\Delta\tilde\eta$ which typically describes a tension exerted on the body. The jump in the extrinsic curvature, instead, adds a finite contribution which is fixed for a given geometry, so that it does not vanish for infinitely close points at different sides of the throat; this implies not a tension but a compression, given that a point at each side of the throat is usually attracted towards it. This particular nature of tides reflects the discontinuous character of the gravitational field (i. e. of the first derivatives of the metric) at {$r_0$}; it suggests that such kind of quite generic wormhole geometry, though traversable in principle, presents the practical problem of great tides acting on a body extended across the throat.

In an analogous way, we can write down the result in \cite{nos21} for the angular tide
of a radially moving object, which is conveniently decomposed as a finite part plus a divergent one:
 \be \label{D_a_perp_gen}
\Delta a_\perp
= \Delta a_\perp^{finite} + \Delta a_\perp^{div},
\ee
where
\begin{eqnarray}
\Delta a_\perp^{finite} & = &
\frac{\Delta x_{\perp}}{2} \,
\left[
{{R^{\varphi 0}}_{\varphi 0}}^-
+\gamma^2 \beta^2 \,
 \left(
 {{R^{\varphi 0 }}_{\varphi 0}}^- -  \, {{R^{\varphi r}}_{ \varphi r}}^-
\right)
\right]_{r_0}\nonumber\\
& & +\,\frac{\Delta x_{\perp}}{2}\left[{{R^{\varphi 0}}_{\varphi 0}}^+ +\gamma^2 \beta^2 \,
 \left(
 {{R^{\varphi 0 }}_{\varphi 0}}^+ -  \, {{R^{\varphi r}}_{ \varphi r}}^+
\right)
\right]_{r_0}
\end{eqnarray}
and introducing the infinitely-short travelling proper time $\delta\tau$ of the object across the shell,
\be
(\Delta a_\perp)^{div}  
=
\Delta x_\perp\, \frac{\gamma \beta \,
{\kappa^{\varphi}}_{\varphi}}{\delta\tau}\,.
\ee
Expressions with the coordinate $z$ instead of $\varphi$ hold for the axial tide in the case of a problem with symmetry along the $z$ axis.
Here, as usual in similar contexts, we  use the velocity $V^{\hat{\mu}} = dx^{\hat{\mu}}/d\tau=(\gamma, \gamma \beta, 0, 0)$ as measured in an orthonormal frame $\{ \vec{e}_{\hat{\mu}}\}$ at rest at the vicinities of the throat to define the parameters $\gamma = 1/\sqrt{1-\beta^2}$ and the positive radial speed $\beta$ of the object as measured in the orthonormal frame. The divergent character of the result is apparent because of the proper time $\delta\tau$, which goes to zero for an infinitely thin shell, in the   denominator. Differing from the case of radial tides, for tides in transverse directions to the radial one both contributions are proportional to the transverse extension $\Delta x_{\perp}$ of the object. For an angular tide, the finite term implies a compression produced by the curvature of the smooth parts of the geometry, while the divergent term corresponds to a stretching effect resulting from the jump in the extrinsic curvature produced by the flare-out at the shell. This serious difficulty could be avoided only in the case of a negligible velocity, or if the geometry is chosen so that it presents no  curvature jump at the throat.

\section{Scalar-tensor theories}

We will go well beyond what could be the analysis of simple corrections to the metric functions (see the Appendix for an example of such kind of approach), by leaving the relativistic framework. We will consider two kinds of scalar--tensor gravity models: the low energy limit of bosonic string theory, yielding the so-called dilaton gravity, and the well known Brans--Dicke theory in which the gravitational constant is replaced by a field whose source is standard matter, and which together with it, determines the spacetime geometry. We will not perform a complete analysis of many cases, but we will restrict to two simple configurations useful to exemplify different aspects of the problem and possible suitable treatments.
The central role in our analysis will be that of the extrinsic curvature tensor: these components are associated to the contribution to the radial tide which does not vanish for infinitely close points, and to the terms in the transverse tides which are formally divergent. Problems in the smooth part can be more easily controlled by suitably choosing certain parameters of the problem (see \cite{nos21}). However we will also examine how the conditions on the extrinsic curvature are reflected in the behaviour of the tides in these regions.

\subsection{Dilaton gravity}

The dilaton spherically symmetric black hole is described by the metric \cite{dil1,dil2,dil3}
\be
ds^2=- f(r)dt^2+ f^{-1}(r)dr^2+ h(r)(d\theta^2+\sin^2\theta d\varphi^2),\label{metric}
\ee
with functions 
\begin{eqnarray}f(r) & = &\lp 1-\frac{A}{r}\rp\lp 1-\frac{B}{r}\rp^{\frac{1-b^2
}{1+b^2}}, \nonumber\\
h(r) & = &r^2\lp 1-\frac{B}{r}\rp^{\frac{2b^2}{1+b^2}}.\label{fyh}
\end{eqnarray}
The constants $A,B$ and the parameter $b$ are related with the mass and electric charge 
of the black hole by
\begin{eqnarray} M & =&\frac{A}{2}+\lp{\frac{1-b^2}{1+b^2}}\rp\frac{B}{2}, 
\nonumber\\
Q&=&\sqrt{\frac{AB}{1+b^2}}.\label{mq}
\end{eqnarray}
The electromagnetic field tensor has non-null 
components $F_{tr}=-F_{rt}=Q/r^{2}$, and the dilaton field is given by 
\be
e^{2\phi}=\lp 1-\frac{B}{r}\rp^{2b/(1+b^2)},
\ee
where the asymptotic value of the dilaton $\phi_{0}$ is taken as zero.  When $b=0$, which corresponds to a 
uniform dilaton, the metric reduces to the  Reissner--Nordstr\"{o}m  geometry, 
while for $b=1$, one obtains $f(r) = 1-2M/r$, $h(r) = r^2\lb 1-Q^2/ (Mr)\rb$.
In what follows, we shall consider $0\le b\le 1$. $B$ and $A$ are, respectively, the inner and outer 
horizons of the black hole; while the outer horizon is a regular event horizon 
for any value of $b$, the inner one is singular for any $b\neq 0$.
If the flare-out condition $h'(r_0)>0$ is fulfilled for $r>r_0$ (with $r_0$ a radius outside the larger horizon surface), two copies of the exterior region can be joined to define a Lorentzian wormhole \cite{dilnos}. 

For such metric, the relevant components of the Riemann tensor are:
\begin{eqnarray}
{R^{r0}}_{r0} & = & -\,\frac{f''(r)}{2}\\
{R^{\varphi 0}}_{\varphi 0} & = & -\,\frac{f'(r)h'(r)}{4h(r)}\\
{R^{\varphi r}}_{\varphi r} & = & {R^{\varphi 0}}_{\varphi 0} -\frac{f(r)}{4h(r)}\left\{ 2h''(r)-\frac{{h'}^2(r)}{h(r)}\right\}
\end{eqnarray}
where we have adopted the notation of a prime for a derivative with respect to the radial coordinate.
The jump of the components of the extrinsic curvature at the shell is given by
\be
{\kappa^\theta}_\theta =   {\kappa^\varphi}_\varphi =\frac{h'(r_0)\sqrt{f(r_0)}}{h(r_0)}
\ee
and
\be
{\kappa^0}_0 = \frac{f'(r_0)}{\sqrt{f(r_0)}}.
\ee

Let us analyze the tides in both the radial and the angular directions at the throat of a thin-shell wormhole. As we noted above, the central problem comes from the extrinsic curvature jump, as the contributions from smooth parts are finite and proportional to the separation between points.  In the radial case this jump leads to a finite, non-vanishing tide between infinitely close points, while in the angular direction the curvature jump implies a divergent tide. In a practical approach, one would always be interested in reducing the corresponding extrinsic curvature jumps ${\kappa^0}_0$ and ${\kappa^\varphi}_\varphi$. Then we will evaluate these elements and examine the behaviour of the quotients ${{\kappa^0}_0}^{Dil}/{{\kappa^0}_0}^{RN}$ and  ${{\kappa^\varphi}_\varphi}^{Dil}/{{\kappa^\varphi}_\varphi}^{RN}$, where ``$Dil$" stands for ``dilaton" and ``$RN$" refers to the pure relativity limit $b\to 0$ corresponding to the Reissner--Nordstr\"{o}m geometry. If we simplify the notation by writing $\frac{1-b^2}{1+b^2}=n$ and $\frac{2b^2}{1+b^2}=m$, for the jump associated to the tide in the radial direction we have
\be
{{\kappa^0}_0}^{Dil} = \frac{f'(r_0)}{\sqrt{f(r_0)}}=\frac{1}{r_0^2}\lp 1-\frac{A}{r_0}\rp^{-\frac{1}{2}}\lp 1-\frac{B}{r_0}\rp^{\frac{n
}{2}-1}\lb A\lp 1-\frac{B}{r_0}\rp+nB\lp 1-\frac{A}{r_0}\rp\rb
\ee
and
\be\label{zeze}
\frac{{{\kappa^0}_0}^{Dil}}{{{\kappa^0}_0}^{RN}}=\lp\frac{r_0-B}{r_0}\rp^{\frac{n}{2}-\frac{1}{2}}\frac{(A+nB) r_0-AB(n+1)}{(A+B)r_0-2AB}.
\ee
To understand this result let us consider two limits: a throat very near the outer horizon, so that $r_0\to A$, and a throat radius very large compared with the horizon radius, for what we take $r_0\to \infty$. For $r_0\to A$,  both in the dilaton as in the pure relativity cases the curvature jump involved in radial tides becomes very large, and we have
\be\label{zeze2}
\frac{{{\kappa^0}_0}^{Dil}}{{{\kappa^0}_0}^{RN}}\to\lp\frac{A-B}{A}\rp^{\frac{n}{2}-\frac{1}{2}}>1
\ee
because $A-B<A$ and for $b>0$ it is $n<1$, and the dilaton would not improve the situation with radial tides; in the other limit, for $r_0\to\infty$ we obtain
\be\label{zeze3}
\frac{{{\kappa^0}_0}^{Dil}}{{{\kappa^0}_0}^{RN}}\to \frac{A+nB}{A+B}<1.
\ee
Therefore the traversability, at least for radially extended objects, is improved in the dilaton framework if the throat is very far from the outer horizon.  Apart from these two limits, one could wonder about the situation for intermediate regions. In the particular  case $b=1$, which corresponds to $n=0$, we have
\be\label{zeze4}
\frac{{{\kappa^0}_0}^{Dil\,(b = 1)}}{{{\kappa^0}_0}^{RN}}= \lp\frac{r_0-B}{r_0}\rp^{1/2}\frac{1}{1+B((r_0-2A)/Ar_0)},
\ee
and it is clear that for any $r_0>2A$ the quotient is smaller than unity. This is a sufficient condition, but not  a necessary one: in fact, both the throat radius and the inner horizon radius could be scaled in terms of the outer horizon by writing $r_0=\alpha_1 A, \ \alpha_1>1$ and $B=\alpha_2 A, \ \alpha_2<1$ to show that the condition to have a better situation with the dilaton field can be written
\be
\frac{{{\kappa^0}_0}^{Dil \,(b = 1)}}{{{\kappa^0}_0}^{RN}}= \lp\frac{\alpha_1-\alpha_2}{\alpha_1}\rp^{1/2}\frac{\alpha_1}{\alpha_1+\alpha_2(\alpha_1-2)}<1.
\ee
Now, one may wonder about the possibility that the improvement of the radial tide across the wormhole throat is achieved at the price of worsening the situation in the smooth regions of the geometry. According to the general analysis of Section 2, answering this question requires to evaluate the behaviour of only one component of the Riemann tensor:  
\be
\frac{{{R^{r0}}_{r0}}^{Dil}}{{{R^{r0}}_{r0}}^{RN}}
=
\frac{(1 - B/r_0)^n r_0 (A B^2 (1 + n) (2 + n) - B (1 + n) (4 A + B n) r_0 + 
   2 (A + B n) r_0^2)}{2 (B - r_0)^2 (-3 A B + (A + B) r_0)}.
\ee
We immediately see that in the convenient limit $r_0 \to \infty$ we have
\be
\frac{{{R^{r0}}_{r0}}^{Dil}}{{{R^{r0}}_{r0}}^{RN}}
\to
\frac{A + nB}{A + B}<1.
\ee
For any $r_0$, in the particular case $b=1\, (n=0)$ considered above we have
\be
\frac{{{R^{r0}}_{r0}}^{Dil (b=1)}}{{{R^{r0}}_{r0}}^{RN}}
=
\frac{A r_0}{B r_0 + A (-3 B + r_0)}
\ee
and we find that for $\alpha_1>3$ we obtain 
\be
\Big| 
\frac{{{R^{r0}}_{r0}}^{Dil (b=1)}}{{{R^{r0}}_{r0}}^{RN}}
\Big|
=
\frac{\alpha_1}{\alpha_1+ \alpha_2(\alpha_1-3)}
<1.
\ee
For these cases the improvement in the curvature jump associated to the radial tide across the throat is then achieved in correspondence with a reduction of the radial tide in the smooth parts of the geometry.

For the jump determining the formally divergent tide in the angular direction we have the analogous expressions
\be
{{\kappa^\varphi}_\varphi}^{Dil} = \frac{h'(r_0)\sqrt{f(r_0)}}{h(r_0)}=\frac{1}{r_0^2}\lp 1-\frac{A}{r_0}\rp^{\frac{1}{2}}\lp 1-\frac{B}{r_0}\rp^{\frac{n
}{2}-1}\lb 2r_0\lp 1-\frac{B}{r_0}\rp+mB\rb
\ee
and
\be\label{fifi}
\frac{{{\kappa^\varphi}_\varphi}^{Dil}}{{{\kappa^\varphi}_\varphi}^{RN}}=\lp\frac{r_0-B}{r_0}\rp^{\frac{n}{2}-\frac{1}{2}}\frac{2(r_0-B)+mB}{2(r_0-B)}.
\ee
Now we can note that for any $n<1$ and $m>0$ both factors in the product above are larger than unity, as long as $r_0$ satisfies the natural condition $r_0> A > B$. Hence no parameter choice in the dilaton model allows to improve the situation with the angular tide in relation with the pure relativistic framework.

\subsection{Brans--Dicke gravity}

As an example within the framework of  Brans--Dicke scalar-tensor gravity we consider the cylindrically symmetric wormholes connecting submanifolds of the form
\begin{equation}
ds^2 = -f(r)dt^2 +g(r)dr^2 +h(r)d\varphi ^2+k(r)dz^2,
\label{e1}
\end{equation}
where $f$, $g$, $h$ and $k$ are positive functions. 
Two copies of the outer region $r\geq r_0$ of this geometry joined at the hypersurface $r=r_0$ constitute a Lorentzian wormhole as long as the flare-out condition is satisfied. For cylindrical configurations we can adopt two definitions of such condition: the usual in which the area of a surface at the throat is minimum, or the circular notion proposed in \cite{brle}, defined by the existence of a minimum of the function $h(r)$ at $r=r_0$ (see also \cite{eisi10}).
 
We will examine the radial and transverse (angular and axial) tides for a radially moving object. The relevant components of the Riemann tensor are
\begin{eqnarray}
{R^{r0}}_{r0} & = & -\,\frac{1}{4f(r)g(r)}\left\{ 2f''(r)-f'(r)\frac{\lp f(r)g(r)\rp'}{f(r)g(r)}\right\},\\
{R^{\varphi 0}}_{\varphi 0} & = & -\,\frac{f'(r)h'(r)}{4f(r)g(r)h(r)},\\
{R^{\varphi r}}_{\varphi r}& = & {R^{\varphi 0}}_{\varphi 0} -\frac{1}{4g(r)h(r)}\left\{ 2h''(r)-h'(r)\frac{\lp f(r)g(r)h(r)\rp'}{f(r)g(r)h(r)}\right\},
\end{eqnarray}
and analogous expressions with $k(r)$ instead of $h(r)$ for the components ${R^{z 0}}_{z 0}$ and ${R^{z r}}_{z r}$. In general there is a matter shell at the throat. The jump of the components of the extrinsic 
curvature across the shell are
\begin{equation}
{\kappa^0}_0= \frac{f'(r_0)
}{f(r_0) \sqrt{g(r_0)}},
\label{e4}
\end{equation}
\begin{equation}
{\kappa^\varphi}_\varphi =  \frac{h'(r_0)}{h
(r_0)\sqrt{g(r_0)}}, 
\label{e5}
\end{equation}
and
\begin{equation}
{\kappa^z}_z= \frac{k'(r_0)}
{k(r_0) \sqrt{g(r_0)}}.
\label{e6}
\end{equation} 

We will carry out a detailed analysis for a class of geometries given by\footnote{To keep dimensions consistent, the radial coordinate is defined as $r = \rho/\rho^*$, where $\rho^*$ can be understood as a ``core'' radius fixing the scale; in what follows we consider $\rho>\rho^*$, then $r \geqslant r_0 >1$, and work with the dimensionless line element $ds^2 = (d\tilde{s}/\rho^*)^2$, where $z \in \mathbb{R}$ and $W^2$ is a dimensionless constant.}
\be
f(r)=g(r)=r^{2d(d-n)+\Om(\om)} ,
\ee 
\be
h(r)=W^2 r^{2(n-d)},
\ee
\be
k(r)=r^{2d}.
\ee
Here $\Om(\om)=[\om(n-
1)+2n](n-1)$, with $\om$ the Brans--Dicke constant (we shall assume $\om>-3/2$ to avoid what could be interpreted as a negative effective gravitational constant; see \cite{grav}) and $d$ 
and $n$ are constants of  integration: $n$ is 
related to the departure from pure general relativity --see 
below-- and $d$ can be understood as a mass parameter; the case $d=0$ within Einstein's gravity describes a conical geometry. The 
Brans--Dicke field has the behaviour 
\be 
\phi=\phi_0r^{1-n}.\label{campo}
\ee
This geometry can be found in Ref. \cite{as00}\footnote{There is a typo in Eq. (31) of that paper, where a factor $(n-1)$ 
multiplying $\om$ is missing, as can be deduced by 
comparison with the immediately preceding equations.}, and can be obtained as the zero current limit of the magnetic solution in Ref. \cite{bd09}; see also \cite{bd15}. The Einstein's gravity 
solutions (see for example \cite{stephani} and also \cite{bronnikov}) are obtained for $n=1$. The usual areal flare-out condition demands $\lp\sqrt{h(r)k(r)}\rp'>0$ at the throat, while the circular condition requires $\lp\sqrt{h(r)}\rp'>0$. In the first case we would say that we have a  throat at $r_0$ as long as $n>0$, while in the second case the inequality $n>d\geq 0$ should hold.  In what follows we will restrict to $0\leq d < 1$ to avoid a circumference decrease for increasing $r$ in the relativistic limit. 

The components of the jump of the extrinsic curvature at the shell take the form
\begin{eqnarray}
{\kappa^0}_0 & = & \frac{2d(d-n)+\Om(\om,n)}{r_0^{d(d-n)+1+\Om(\om,n)/2}},\\
{\kappa^\varphi}_\varphi & = & \frac{2(n-d)}{r_0^{d(d-n)+1+\Om(\om,n)/2}},\\
{\kappa^z}_z & = &
 \frac{2 d}{r_0^{d(d-n)+1+\Om(\om,n)/2}}.
 \end{eqnarray}
The relativity limit $n=1$ reduces to
\begin{eqnarray}
{\kappa^0}_0 & = & \frac{2d(d-1)}{r_0^{d(d-1)+1}},\\
{\kappa^\varphi}_\varphi & = & \frac{2(1-d)}{r_0^{d(d-1)+1}},\\
{\kappa^z}_z & = &
 \frac{2 d}{r_0^{d(d-1)+1}}.
\end{eqnarray}
Besides, the jump of the scalar field across the surface given by  $\langle\phi_{,N}\rangle = 8\pi S/(3+2\omega)$, where $S$ is the trace of the surface energy-momentum tensor of the shell, imposes the condition \cite{eisibd}
\be
2\om\frac{\phi'(r_0)}{\phi(r_0)}=\frac{f'(r_0)}{f(r_0)}+\frac{h'(r_0)}{h(r_0)}+\frac{k'(r_0)}{k(r_0)},\label{phiprima}
\ee
which implies a constraint involving the throat radius $r_0$ and all the parameters that must be fulfilled  by the wormhole construction from a given metric\footnote{The jump of the normal derivative of $\phi$ comes from the fact that the second derivatives of the field are proportional to the trace of the energy-momentum tensor, which is singular at $r=r_0$. This is not the case for the dilaton field, whose second derivatives are proportional to the trace of the electromagnetic field, so no singular contribution appears associated to the existence of an infinitely thin matter layer. Therefore, the normal derivative of the dilaton field must be continuous across the shell; this is automatically satisfied, and no restrictions exist on the  parameters of the configurations.}. In this particular example the condition turns to give only the relation 
\be
\om(1-n^2)=2(d^2+n^2-nd)
\ee 
which can be used to write, for example, $\om$ in terms of the parameters $n$ and $d$. We then obtain
\begin{eqnarray}
{\kappa^0}_0 & = & \frac{4d(d-n)+2n(n-1)}{(n+1)r_0^{1+\lp 2d(d-n)+n(n-1)\rp/(n+1)}},\\
{\kappa^\varphi}_\varphi & = & \frac{2(n-d)}{r_0^{1+\lp 2d(d-n)+n(n-1)\rp/(n+1)}},\\
{\kappa^z}_z & = & \frac{2d}{r_0^{1+\lp 2d(d-n)+n(n-1)\rp/(n+1)}}.
\end{eqnarray}
Note that for $d=0$, which in the pure relativistic framework would correspond to a wormhole connecting two conical geometries, the component ${\kappa^z}_z $ determining the axial tide vanishes. 

Now the question would be how does each component behave with the parameter $n$ determining the departure from pure relativity. Does $n\neq 1$ reduce the curvature jump? The answer will, in principle, depend on the throat radius $r_0$. A  straightforward way to perform the analysis could be to take ${\kappa^0}_0$, ${\kappa^\varphi}_\varphi$ and ${\kappa^z}_z$ as functions of the parameter $n$ and study the derivatives at $n=1$: the relativity (Rel) limit would give the best conditions for traversability only if for $n=1$ we find a radius $r_0$ such that there is a minimum of the elements of the curvature jump. If this is not the case, the sign of the first derivative of each component will indicate whether we must consider a choice of $n<1$ or $n>1$ within the Brans--Dicke (BD) gravity framework. This also seems to be the more suitable approach for more general geometries in the same framework; however, for this example we can obtain a direct first insight in the dependence with $n$ by writing quotients in the same way of the preceding section:
\be
\frac{{{\kappa^0}_0}^{BD}}{{{\kappa^0}_0}^{Rel}}=\frac{\lb 2d(d-n)+n(n-1)\rb r_0^{(d(n-1)(d+1)-n(n-1))/(n+1)}}{d(d-1)(n+1)},
\ee
\be
\frac{{{\kappa^\varphi}_\varphi}^{BD}}{{{\kappa^\varphi}_\varphi}^{Rel}}=\frac{(d-n)r_0^{(d(n-1)(d+1)-n(n-1))/(n+1)}}{(d-1)},\\
\ee
\be
\frac{{{\kappa^z}_z}^{BD}}{{{\kappa^z}_z}^{Rel}}
= 
r_0^{
(d (n - 1) (d + 1) - n (n - 1))
/(n+1)}.
\ee
We immediately see that for large $n$ the quotient of radial tides behaves in the form 
$\sim- nr_0^{-n}$, 
for the angular tide it is of the form 
$\sim nr_0^{-n}$,
while for the axial tide it is of the form 
$\sim r_0^{-n}$.
Because $r_0>1$, the ratios vanish in the limit $n\to\infty$ being the axial case the faster one.
Hence, for all possible orientations, the tides are weaker in the case of a very large departure from pure relativity; the Brans--Dicke framework thus constitutes a better situation in what regards traversing the throat  of this kind of wormhole geometry.
The behaviour of the radial tide for a radially moving object in the smooth regions of the spacetime is inferred from the quotient
\be
\frac{{{R^{r0}}_{r0}}^{BD}}{{{R^{r0}}_{r0}}^{Rel}}
=
\frac{
(2 d^2 - 2 d n + ( n-1) n) r_0^{2(d (n - 1) (d + 1) - n (n - 1))
/(n+1)}
}{
(d-1) d (n+1)
}.
\ee
For large $n$ this behaves as $\sim- nr_0^{-2n}$, so it vanishes for $n\to\infty$. The dependence of the angular tide in any of the two smooth submanifolds joined at the throat can be described by the quotient of the finite parts:
\begin{eqnarray}
\frac{\Delta a_{\perp(\varphi)}^{finite}|^{BD}}{\Delta a_{\perp(\varphi)}^{finite}|^{Rel}}
& = &
\frac{(d - n)
(d^2 (2 + 4 \gamma\beta) + ( n-1) (n + ( n-1) \gamma\beta) + 
   d (\gamma\beta- n (2 + 3 \gamma\beta)))}{(d-1) d (1 + n) ( 
   d-1 - \gamma\beta + 2 d \gamma\beta)}\nonumber\\
   & & \times\  r_0^{2(d (n - 1) (d + 1) - n (n - 1))
/(n+1)},
\end{eqnarray}
with the velocity dependent parameters $\beta$ and $\gamma$ defined as above. The large $n$ behaviour has the form $\sim n^2 r_0^{-2n}$, and vanishes in the limit $n\to\infty$.
For the tide along the symmetry axis we have
\begin{eqnarray}
\frac{\Delta a_{\perp(z)}^{finite}|^{BD}}{\Delta a_{\perp(z)}^{finite}|^{Rel}}
& = &
\frac{(
\gamma\beta+ d^2 (2 + 4 \gamma\beta) + 
 n (n -1- \gamma\beta + 2 n \gamma\beta) - 
 d ( \gamma\beta + n (2 + 5 \gamma\beta))
)}
{
( d-1) (n+1) (d -  \gamma\beta+ 2 d \gamma\beta)}\nonumber\\
& & \times \ r_0^{2(d (n - 1) (d + 1) - n (n - 1))
/(n+1)}.\end{eqnarray}
For large $n$ we find a behaviour $\sim n r_0^{-2n}$.
Summarizing: for this example, in Brans--Dicke gravity  the reduction of tides in the smooth regions of the geometry is in correspondence with the improvement of tides across the throat.

\section{Discussion}
If taken literally, previous results \cite{nos21} about tides across the throat of thin-shell wormholes would rule out a safe travel from one side to the other: as a result of the jump of the extrinsic curvature at the throat, angular tides are formally divergent, and radial tides lead to a finite but non vanishing relative acceleration between two infinitely close points. However, an alternative more realistic interpretation of the formal results as an approximation to wormholes supported by exotic matter shells of little but non vanishing thickness is a reasonable possibility. Such point of view naturally leads to the problem of determining the conditions to reduce the jump of the components of the extrinsic curvature across the shell. Here we have examined the consequences of extending the traversability analysis to frameworks beyond relativity, as the scalar-tensor theories  known as dilaton gravity and Brans--Dicke gravity. We have studied particular examples with spherical symmetry for the first, and with cylindrical symmetry for the second. We have considered certain limits and shown that in the dilaton example no general improvement is possible, as the dependence of angular tides with the departure from pure relativity is opposite to that of radial tides; however, in the case in which a reduction of the extrinsic curvature jump across the throat is possible, the suitable configurations also reduce the  corresponding tide in the smooth parts of the geometry.  In the Brans--Dicke cylindrical example, instead, a large departure from relativity seems to improve the situation with both transverse (i.e. angular and axial) and radial tides; moreover, this is in correspondence with a reduction of tides in the smooth regions of the geometry. In the appendices we have also carried out a limited analysis of other examples within the relativistic framework. We have considered the behaviour of tides in relation with certain parameters associated to the sources determining the geometries, both from the point of view of the extrinsic curvature jump, as also in an approximate more elementary study of the radial relative acceleration between points at different sides of the wormhole throat. 

\section*{Appendix A: Corrections in spherical symmetry}

Let us discuss a very simple consideration about the consequences of additional terms in the metric coefficients of geometries with a given symmetry. Take the spherically symmetric geometry
\be
ds^2=-f(r)dt^2+f^{-1}(r)dr^2+r^2\left(d\theta^2+\sin^2\theta d\varphi^2\right)
\ee
so that from two copies of it we construct a symmetric wormhole with extrinsic curvature jump at the throat given by
\be
{\kappa^\varphi}_\varphi =\frac{2}{r_0}\sqrt{f(r_0)}
\ee
and
\be
{\kappa^0}_0 = \frac{f'(r_0)}{\sqrt{f(r_0)}}.
\ee
If we introduce  a correction to the geometry changing $f(r)$ to $f^*(r)=f(r)+q(r)$, we will have the new jumps
\be
{{\kappa^\varphi}_\varphi}^* =\frac{2}{r_0}\sqrt{f(r_0)+q(r_0)}
\ee
and
\be
{{\kappa^0}_0}^* = \frac{f'(r_0)+q'(r_0)}{\sqrt{f(r_0)+q(r_0)}}.
\ee
From this we immediately see that a negative $q(r)$ at $r_0$ improves the situation with the angular tide, while  a positive addition to the original metric makes things worse in this direction. However, with the component ${\kappa^0}_0$ determining the radial tide the situation presents more possibilities: the case  $q(r_0)>0$ and $q'(r_0)<0$ diminishes the radial tide, while the opposite happens for the reverse signs, and a particular analysis is required by the other two cases. A clear example is provided by the Reissner--Nordstr\"om (RN) wormhole \cite{ern}, with the geometry at each side understood as a correction $Q^2/r^2$ added to the Schwarzschild (Sch) metric function $f(r)=1-2M/r$. We have
\be
{{\kappa^\varphi}_\varphi}^{RN} =\frac{2}{r_0}\sqrt{f(r_0)}=\frac{2}{r_0}\sqrt{1-\frac{2M}{r_0}+\frac{Q^2}{r_0^2}},
\ee
and
\be
{{\kappa^0}_0}^{RN} = \frac{f'(r_0)}{\sqrt{f(r_0)}}=\frac{2}{r_0^2}\lp\frac{Mr_0-Q^2}{\sqrt{r_0^2-2Mr_0+Q^2}}\rp.
\ee 
A direct inspection shows that the Schwarzschild case $(Q=0)$ gives a smaller jump ${\kappa^\varphi}_\varphi$ to which the angular tide is proportional. But the relation is reversed for ${{\kappa^0}_0}$, which is larger for $Q=0$. Hence the addition of the charge improves the situation for the radial tide, but makes things worse for the angular tide. 
Note that, for a certain throat radius $r_0$, a decrease of ${\kappa^\varphi}_\varphi$ must necessarily come from a smaller value of $f(r_0)$, which in turn implies an increase of ${\kappa^0}_0$; this can only be avoided by an appropriate decrease of $f'(r_0)$. In general, in the line of Section 3.2, we can pose the problem as follows: write the metric coefficient as a function of $r_0$
and a parameter $p$, and study the behaviour of  the derivatives $\partial {\kappa^\varphi}_\varphi(r_0,p)/\partial p$ and  $\partial {\kappa^0}_0(r_0,p)/\partial p$; the best possible situation will be given in case that a parameter $p$ exists such that both derivatives have the same sign for $r_0$ an admissible throat radius. Then a  suitable choice of $p$ allows to reduce tides in both the radial and the angular directions. When such possibility
does not exist, as with the charge in the Reissner--Nordstr\"om example here, one must select which is the most convenient  direction to reduce the corresponding tide. Of course, this simple considerations can easily be extended to more general metrics and to more than one parameter.

\section*{Appendix B: A fine tuning approach}

The nature of the problems manifest in the infinitely thin shell limit suggests a somewhat different analysis; the point is that instead of assumptions about the dimensions of shells involved, or the search for  ways to reduce the extrinsic curvature jumps, for negligible speeds we can try a fine tuning which, of course, is not always desirable as a rule. For radial tides on rest or moving objects the central point in our traversability analysis is the necessity of a surmountable
relative acceleration between two infinitely close points separated by the wormhole throat. While this is automatically achieved in locally flat geometries or in those spacetimes with constant $g_{00}$ and, thus, null acceleration, another possibility could be to admit two geometries connected at a radius $r_0$ such that just at each side we have a locally vanishing acceleration. The condition for a null acceleration at each side implies $g'_{00}(r_0)=0$. The fact  that near $r_0$ no problems arise associated to jumps of the acceleration is ensured by the assumption that the submanifolds joined at the wormhole throat are well behaved; as a practical rule, 
in order to avoid large, though smooth, changes in $g'_{00}$ in the same vicinity,
we could include the condition of a small second derivative $g''_{00}$. Of course, nothing in these considerations points to solving or at least improving the difficulties with angular tides; at this point we should just rely on a quasi static crossing through the throat.

Let us analyze some examples. In spherically symmetric backgrounds, such behaviour of the metric is not possible, for example, for finite radii in a Schwarzschild geometry. But consider, instead, the thin-shell wormhole connecting two Reissner--Nordstr\"om geometries \cite{ern}. At each side of the throat the metric has the form 
\be
ds^2=-f(r)dt^2+f^{-1}(r)dr^2+r^2\left(d\theta^2+\sin^2\theta d\varphi^2\right)
\ee
with 
\be
f(r)=1-\frac{2M}{r}+ \frac{Q^2}{r^2},
\ee
where $Q$ is the electric charge. The condition $g'_{00}(r_0)=-f'(r_0)=0$ is fulfilled at 
\be
r_0=\frac{Q^2}{M};
\ee
then a wormhole with such throat radius would present no tidal problems, at least at the minimal area surface. Note that if we insist in $|Q|<M$, which is usually required to avoid a naked singularity in the complete Reissner--Nordstr\"om geometry, this is not compatible with the condition of $r_0$ greater than the event horizon radius $r_h=M+\sqrt{M^2-Q^2}$. Hence we should start this wormhole construction from two geometries including a naked singularity; however, this is not a problem within the context in which the regions $r<r_0$ are removed in the cut and paste mathematical construction. 

In cylindrically symmetric backgrounds we have a similar situation: while the power-law behaviour of, for example, the well known Levi--Civita metric excludes the possibility to achieve the condition $g'_{00}=0$, more general axisymmetric wormholes as those associated to Einstein--Maxwell spacetimes (see \cite{eisi10} and also \cite{stephani,bronnikov}) could provide the freedom necessary to fulfil the conditions required, for certain values of the parameters.  The pure electric case corresponding to a constant charge distribution along the symmetry axis and a radial electric field does not allow for $g'_{00}=0$. However, let us consider the geometry associated to a constant current along the $z$ axis giving an angular magnetic field, that is 
\be
ds^2= r^{2m^2}G^2(r)(-dt^2+dr^2)+r^2G^2(r)d\varphi^2+G^{-2}dz^2
\ee
with 
\be
G(r)=k_1r^m+k_2r^{-m}
\ee
and $m, k_1, k_2$ constants such that $k_1k_2>0$. This metric admits $g'_{00}=0$ at a radius $r_0$ satisfying 
\be
r_0^{2m}=\frac{k_2(1-mr_0)}{k_1(1+mr_0)};
\ee
given the requirement $k_1k_2>0$, then there are two possibilities: $m<0,\ -mr_0<1$ and $m>0,\ mr_0<1$. If two copies of the outer part of such spacetime are joined at $r_0$, then the resulting thin-shell wormhole would have no traversability problems coming from {radial} strong tides at the throat. In general, one should also verify the flare-out condition at the radius $r_0$ (that is, that we have a minimal area {per unit length} or minimal radius surface at $r=r_0$); in this pure magnetic case the mostly adopted areal version of this condition is automatically fulfilled, as a direct inspection shows. Note, however, that besides the fact that the angular tides are not solved, differing from the gauge cosmic string example where $g_{zz}=1$ ensured the absence of problems associated with tides in the direction parallel to the axis, here we have $g_{zz}=G^{-2}(r)\neq 1$; this implies a new difficulty which cannot be simultaneously avoided. In fact, we must choose between two directions which will be the safe one: to avoid dangerous tensions or pressures along the $z$ axis we should  demand $g'_{zz}(r^*_0)=0$, which yields $r^*_0=\left(k_2/k_1\right)^{1/(2m)}$ 

%These considerations imply a fine tuning which may be not desirable in general. Nevertheless, it could be right if we understand it as an operational simplification associated to a more realistic condition, i.e. the requirement of a sufficiently small acceleration $a$ at each side of a shell of finite though little thickness $\epsilon$, so that the quotient $\Delta a/\epsilon$ is admissible. 
%This approximation would be in the line of the ``traversability in practice'' condition adopted in, for instance, Ref. \cite{book} for wormholes which are not of the thin-shell class, where a maximum quotient $g/l$ is admitted if an object can withstand a maximum tidal acceleration $g$ between two points separated by a distance $l$. In a sense, the analysis of the next section follows this line, as we will understand the formal results as a guide to find conditions improving the problems noted above.


\vspace{-0.3cm}
\begin{thebibliography}{99} 

\bibitem{nos21} E. Rub\ai n de Celis and C. Simeone, Eur. Phys. J. C \textbf{81}, 937 (2021).

\bibitem{martin} M. G. Richarte and C. Simeone, Int. J. Mod. Phys. D {\bf  17}, 1179 (2008).

\bibitem{book} M. Visser, \textit{Lorentzian Wormholes} (AIP Press, New York, 1996).

\bibitem{grav} S. Weinberg, {\it Gravitation and Cosmology}, John Wiley and sons, New York (1972).

\bibitem{dil1} D. Garfinkle, G. T. Horowitz and A. Strominger, Phys. Rev. D \textbf{43}, 3140 (1991); ibid. \textbf{45}, 3888(E) (1992).

\bibitem{dil2} J. H. Horne and G. T. Horowitz,  Phys. Rev. D \textbf{46}, 1340 (1992).

\bibitem{dil3} G. W. Gibbons and K. Maeda, Nucl. Phys. B \textbf{298}, 741 (1988).

\bibitem{dilnos} E. F. Eiroa and C. Simeone, Phys. Rev. D \textbf{71}, 127501 (2005).

\bibitem{brle} K. A. Bronnikov and J. P. S. Lemos, Phys. 
Rev. D {\bf 79}, 104019 (2009).

\bibitem{eisi10} E. F. Eiroa and  C. Simeone, Phys. Rev. D \textbf{81}, 084022 (2010); Erratum: Phys. Rev. D \textbf{90}, 089906 (2014).

\bibitem{as00} A. Arazi and C.  Simeone,  Gen. Relativ. 
Gravit. {\bf 32}, 2259 (2000).

\bibitem{bd09} A. Baykal and O. Delice, Gen. Relativ. 
Gravit. {\bf 41}, 267 (2009).
   
\bibitem{bd15} P. Kirezli and O. Delice, Phys. Rev. D \textbf{92}, 104045   (2015).

\bibitem{stephani} H. Stephani, D. Kramer, M. MacCallum, C. Hoenselaers, E. Herlt, Exact Solutions of Einstein’s
Field Equations, Cambridge University Press (2003)

\bibitem{bronnikov} K. Bronnikov, N. O. Santos, Anzhong Wang, Class. Quantum Grav. \textbf{37}, 113002 (2020).

\bibitem{eisibd} E. F. Eiroa and C. Simeone, Phys. Rev. D \textbf{82}, 084039 (2010);  arXiv:1008.0382v3 [gr-qc] (2016).

\bibitem{ern} E. F. Eiroa and G. E. Romero, Gen. Rel. Grav. \textbf{36}, 651 (2004).

\end{thebibliography}
\end{document}